\documentclass[10pt]{article}
\usepackage{authblk}
\usepackage{amsmath}
\usepackage{amsfonts}
\usepackage{amssymb}
\usepackage{graphicx,xcolor}
\usepackage{xpatch}

\usepackage[numbers,comma,sort&compress]{natbib}

\usepackage{fullpage}

\newcommand{\Bn}{B_n}
\newcommand{\Bt}{B_t}
\newcommand{\str}{\varepsilon}
\newcommand{\shat}{\hat{\str}}
\newcommand{\what}{\hat{w}}
\newcommand{\vhat}{\hat{v}}
\newcommand{\fhat}{\hat{f}}

\newcommand{\za}{z}
\newcommand{\zb}{\bar{z}}
\newcommand{\bT}{\boldsymbol{T}}
\graphicspath{{./Figures/}}

\title{ \bf Curvature-induced stiffening of a fish fin}
\date{}
 
\author[1]{Khoi Nguyen}
\author[2,**]{Ning Yu}
\author[1]{Mahesh M. Bandi}
\author[3]{Madhusudhan Venkadesan}
\author[4,*]{Shreyas Mandre}
\affil[1]{Collective Interactions Unit, OIST Graduate University, Onna, Okinawa, 904-0495, Japan}
\affil[2]{Department of Mechanical Engineering, Tsinghua University, Beijing, 100084, P. R. China}
\affil[3]{Department of Mechanical Engineering and Materials Science, Yale University, New Haven, CT 06520, USA}
\affil[4]{School of Engineering, Brown University, Providence, RI 02912, USA}
\affil[*]{corresponding author: shreyas\_mandre@brown.edu}
\affil[**]{research performed while visiting Brown University}

\begin{document}
\maketitle
\begin{abstract} 
How fish modulate their fin stiffness during locomotive manoeuvres remains unknown.
We show that changing the fin's curvature modulates its stiffness.
Modelling the fin as bendable bony rays held together by a membrane, we deduce that fin curvature is manifested as a misalignment of the principal bending axes between neighbouring rays. 
An external force causes neighbouring rays to bend and splay apart, and thus stretches the membrane.
This coupling between bending the rays and stretching the membrane underlies the increase in stiffness.
Using 3D reconstruction of a Mackerel ({\it Scomber japonicus}) pectoral fin for illustration, we calculate the range of stiffnesses this fin is expected to span by changing curvature.
The 3D reconstruction shows that, even in its geometrically flat state, a functional curvature is embedded within the fin microstructure owing to the morphology of individual rays.
Since the ability of a propulsive surface to transmit force to the surrounding fluid is limited by its stiffness, the fin curvature controls the coupling between the fish and its surrounding fluid.
Thereby, our results provide mechanical underpinnings and morphological predictions for the hypothesis that the spanned range of fin stiffnesses correlates with the behaviour and the ecological niche of the fish.
\end{abstract}

\thispagestyle{empty}
\section{Introduction}

Ray-finned fish, or Actinopterygii, use their fins as a general purpose device to manipulate the surrounding fluid.
That the rayed structure of the fin is found in greater than 99\% of all living fish species is a testament to its versatility in the aquatic environment \cite{lund1967analysis,gosline1971functional,lauder1989caudal}.
The fin is used in a wide range of manoeuvres \cite{sfakiotakis1999review}, from holding stationary in spatio-temporally varying turbulent flows to explosive bursts of motion.
Insight into design principles underlying its biomechanics could therefore shed light into its overwhelming dominance in the aquatic realm, and find applications in underwater robotics. 

The mechanical stiffness of the fin is a key parameter that influences its propulsive performance \cite{alben2008optimal,michelin2009resonance,spagnolie2010surprising,park2012kinematic,esposito2012robotic,rosic2017performance}.
What structural elements might be responsible for the fin's stiffness?
The fin stiffness under bending deformations is tacitly assumed to equal the individual ray bending rigidity multiplied by the number of rays \cite{mccutchen1970trout}.
Experimentally, an individual ray's bending rigidity is estimated by supporting it in a cantilevered loading and measuring the displacement at a point where a force is applied to deform the ray \cite{alben2007mechanics,flammang2013functional}.
Here we show using a mathematical model that fin curvature along a direction transverse to the rays stiffens it to bending beyond this simple picture.
The mechanism of this stiffening underlies the common observation that, under its own weight, a flat sheet of paper droops but a slightly curved sheet stiffens and barely deforms. 
Curvature couples out-of-plane bending of the sheet to its in-plane stretching; the in-plane elastic properties of the sheet thus become relevant in determining its stiffness.
However, unlike a sheet of paper, rayed fins are composite structures consisting of hard bony rays interconnected by softer membranes \cite{westneat2004structure,lauder2004morphology}.
We develop a simple mathematical model that encapsulates the essence of the bending-stretching coupling applied to the rayed-fin structure; bending of the rays is mechanically coupled to the stretching of the membranes.
Via this mechanism, the fish can modulate the stiffness of the fin by merely changing its transverse curvature.
We calculate the range of stiffnesses spanned by a model fin as a function of curvature and find the important underlying parameters to be the anisotropy in the bending modulus of the rays and the elastic properties of the membrane under stretching.
Finally, an important result from our model is the definition of the discrete curvature itself.
We identify it to be proportional to the misalignment of the principal axis of bending.
By applying this definition, we identify a ``functional'' curvature although the fin may be geometrically flat. 
We apply our model to a sample fish fin -- a Mackerel ({\it Scomber japonicus}) pectoral fin -- in which we predict the degree of curvature-induced stiffening and illustrate the presence of a functional curvature.
We emphasize that the Mackerel fin was not chosen for a biological function it specializes in.
It merely serves to illustrate the mechanical principles, which are the focus of this article.

\begin{figure}[t]
{\includegraphics[width=0.98\textwidth]{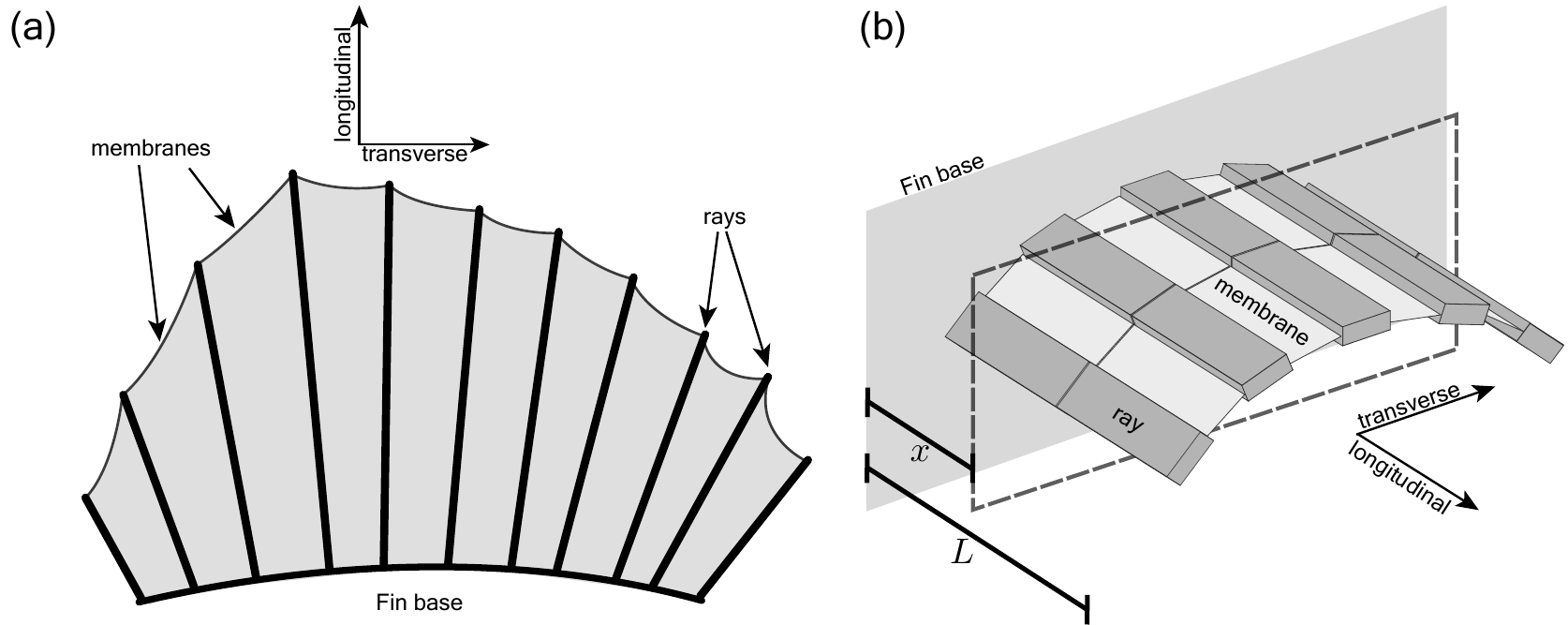}}
\caption{Schematics for a model fish fin. 
(a) A prototypical fish fin has multiple bony rays which attach to a common fin base and are held together by tissue membranes.
(b) We model the fin using elastic beams for rays and distributed springs for membranes.
The rays and membranes are of length $L$ and approximate a transverse curvature by tracing a constant circular arc.
}
\label{fig:FinSchematic}
\end{figure}

\section{Mathematical model} 
\label{sec:mathmodel}
A fish fin has many features that complicate its elastic behaviour.
To list a few, individual rays have non-uniform length and heterogeneous internal structures, and the tissue membranes have nonlinear elasticity (see figure \ref{fig:FinSchematic}(a)).
To understand the underlying mechanics and isolate the essential principle of bending-stretching coupling, we make the following simplifying assumptions.
The membranes are assumed to be identical in composition and dimensions, and resist stretching transverse to the rays with a distributed elastic force.
At every distance $x$ from the fin base, the distributed force equals the amount of stretching multiplied by a spring constant $k$.
The fin rays are approximated as linearised Euler-Bernoulli beams with bending rigidities $\Bn$ and $\Bt$ along principal axes that are aligned with the normal and tangent to the fin surface, respectively.
The material of the fin rays and membrane is assumed to be linearly elastic, i.e., the local strain in the material is linearly proportional to the local stress.
Because of the transverse curvature, the principal bending directions of each ray is misaligned with its neighbours.
A schematic of the fin is shown in figure \ref{fig:FinSchematic}(b) where bending-stretching coupling acts parallel to the cross-section shown (dashed box).
With these assumptions, we construct a fin model with few parameters, yet maintain a meaningful approximation to the curvature-induced stiffening of rayed fins.

\begin{figure}[t]
{\includegraphics[width=0.98\textwidth]{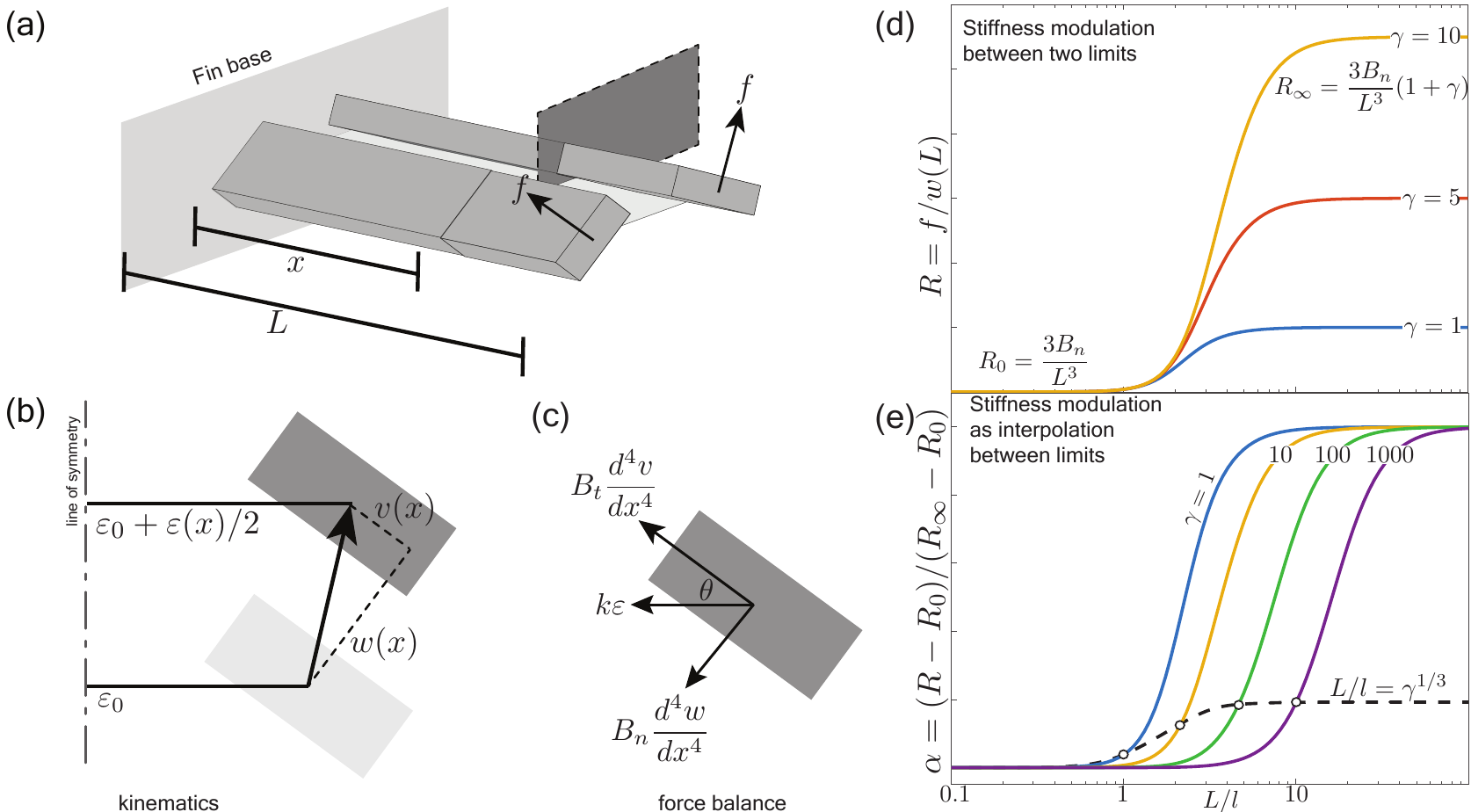}}
\caption{Results from the analysis of stiffening by curvature in a 2-ray fin. 
(a) Two rays that are symmetric about a membrane. The rays deform under point forces $f$ applied at the ray tips along their soft principal bending directions. 
(b) A cross-sectional view of the 2-ray fin at a distance $x$ from the fin base.
The displacement of a ray due to point force loadings is decomposed into its principal components, $w(x)$ and $v(x)$.
The membrane stretches by $\str(x)$ due to the displacement.
(c) Forces acting on the ray.
$\theta$ is the angle the principal axes form with the membrane, $\Bt$ and $\Bn$ are the tangential and normal bending rigidities respectively, and $k$ is the membrane spring constant.
(d) Fin stiffness $R$ transitions between two asymptotic limits as a function of the coupling parameter $L/l$. The larger the value of $\gamma$, the greater the maximal stiffness that can be achieved by an infinitely stiff membrane ($L/l \to \infty$).
(e) Rescaling stiffness as an interpolation shows that $\gamma$ also acts to delay the transition to stronger coupling, which is characterized by a dashed black line.
}
\label{fig:cs}
\end{figure}
\subsection{Bending-stretching coupling in a 2-ray fin}
\begin{figure}[t]
{\includegraphics[width=0.98\textwidth]{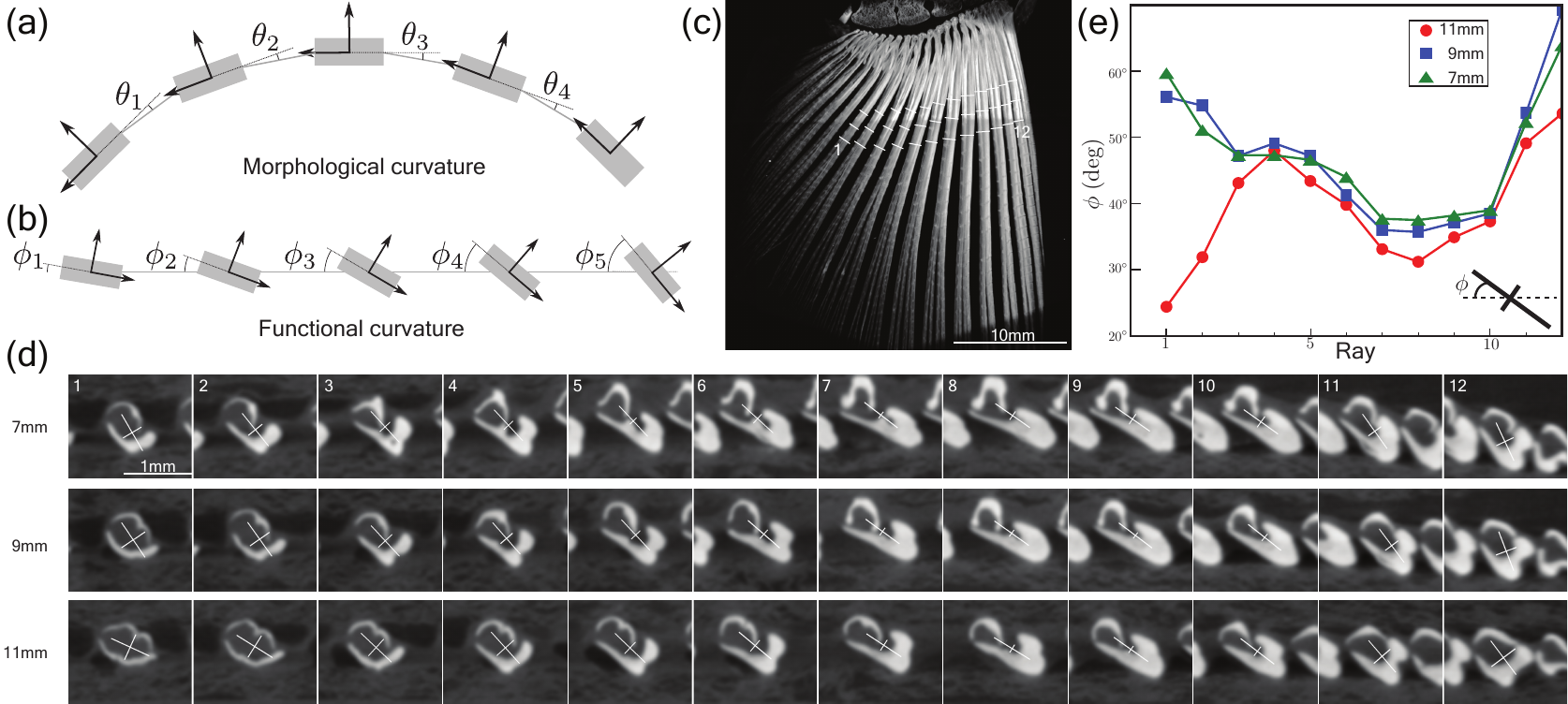}}
\caption{
Morphological and functional curvature in rayed-fin model.
(a) Schematic of a multiray fin with morphological curvature. The curvature may be represented by the misalignments of the principal bending directions of rays. In this case, the curvature arises because the fin as a whole is bent in a curved shape transverse to the rays.
(b) Schematic of a multiray fin with functional curvature. While the fin is morphologically flat, the misalignment between the principal axes of local fin bending moments causes out-of-plane bending to be coupled with stretching of the membrane, and therefore characterizes the functional curvature.
(c) Planform perspective of the three-dimensional reconstruction of an isolated Mackerel  pectoral fin using $\mu$CT.
The Mackerel fin is stretched by extending the first and last rays apart, and held flat between two compressive plates with soft contact surfaces.
The bony rays appear bright, whereas the soft elastic membranes appear dark.
(d) Representative cross-sections of rays taken at fixed distances of 7, 9, and 11 mm from the fin base and taken normal to each ray's central axis. The cross-sections are depicted in (c) as white cut-lines. 
The cross-hairs superimposed on the image depicts the principal axes and magnitudes of the anisotropic second moment of the bright regions, and represents the local bending rigidity of the rays.
(e) The angles made by the principal axes of each ray with the plane of the fin (and the membrane) for three different fixed distances. 
The difference in angles between any two adjacent rays allows for bending-stretching coupling and characterizes the functional curvature of this sample fin.
The lower angles for the first three rays at 11mm are likely due to the onset of distal branching.
}
\label{fig:Springs}
\end{figure}

The basic unit of bending-stretching coupling is a pair of rays connected by a single membrane, a 2-ray fin (figure~\ref{fig:cs}a).
A fin with multiple (say $N$) rays is simply a collective set of 2-ray units, and an exact account of the analogy between a $N$-ray and a 2-ray fin is presented in Appendix A.
Figures~\ref{fig:cs}(a--c) depict the kinematics and balance of forces for a 2-ray fin at a distance $x$ from the fin's base. 
Each ray has a hard and a soft principal bending direction, depicted as an anisotropic rectangular cross-section.
The misalignment $\theta$ between the principal directions and the membranes encodes the fin's transverse curvature, and defined such that the hard and soft directions are tangential and normal to the fin surface, respectively.
That is to say, the membrane surface differs from the fin surface, and this becomes apparent in the $N$-ray fin of figure \ref{fig:Springs}(a) where the fin surface at any particular ray is the average orientation of its two neighbouring membranes.

The 2-ray fin deforms due to an externally applied force causing the rays to bend with displacements $w(x)$ and $v(x)$ along the soft and hard bending directions, respectively. 
In turn, this causes the membrane to elongate by $\str(x)$ and resist further displacement of the rays, thereby stiffening the fin via bending-stretching coupling.

To quantify the stiffness, we cantilever the model fin at its base, and apply a point force $f$ at the ray tips $x=L$ along their normal, soft direction.
The stiffness $R$ of the fin is defined as the ratio of the applied force to the normal displacement of the tip which manifests as $w(L)$ in the 2-ray unit,
\begin{align}
 R = \dfrac{f}{w(L)},
\end{align}
in the limit of small displacements. 
We determine the influence of fin curvature and membrane elasticity on $R$ by solving the mathematical model we develop here.
In the absence of the membrane and the bending-stretching coupling, the stiffness $R$ in our model will arise simply due to the bending rigidity of the rays.
Our loading scheme and stiffness definition thus allows us to determine whether the membrane, when present, also plays a role.
The external point load in our model is applied on the rays instead of the membranes, accounting for the transmission of the hydrodynamic force on the membranes to the rays at the point of attachment, as described in Appendix B.
The rationale for ignoring the bulging deformation of the membrane due to hydrodynamic loading is also developed in the Appendix B.
Note that limiting the analysis to infinitesimal displacements implies that the ray curvature along its length changes infinitesimally due to bending, and does not influence the simplified analysis we present next.


The coupling between ray bending and membrane stretching because of fin curvature appears in a geometric relation between the displacements as
\begin{align}
 \str = 2w \sin\theta  + 2v \cos\theta, \label{eqn:2raygeo}
\end{align}
and in the balance of forces acting on the rays,
\begin{align}
 \Bn w'''' = -k \str\sin\theta, \quad \Bt v'''' = -k \str\cos\theta, \label{eqn:2ray}
\end{align}
where primes denote derivatives with respect to $x$.
In the force balance \eqref{eqn:2ray}, $k\str$ is a distributed elastic force that resists membrane stretching, whereas $\Bn w''''$ and $\Bt v''''$ are the ray resistances to bending in the normal and tangential directions, respectively.
The rays are clamped at the base, and loaded at their tips in the normal direction as given by
\begin{gather}
\begin{aligned}
&w = w' = v = v' = 0							 &\text{ at } x = 0\ \\
&w'' = v'' = \Bt v''' = 0, \quad \Bn w''' = -f    &\text{ at } x = L.\\
\end{aligned}
\label{eqn:bc}
\end{gather}

A physical length scale $l$ arises from the competition between ray bending and membrane stretching. It is found by eliminating $v$ and $w$ from equations (\ref{eqn:2raygeo}-\ref{eqn:2ray}) and examining the behaviour of membrane elongation $\str$, which satisfies 
\begin{gather}
\str'''' = -\frac{1}{l^{4}}\str, \quad \text{ where } \quad \frac{1}{l^4} = \frac{2k\cos^2\theta}{\Bt} \left( 1 + \gamma \right), \qquad 
\gamma = \dfrac{\Bt}{\Bn}\dfrac{\sin^2\theta}{\cos^2\theta}. \label{eqn:2raystr}
\end{gather}

The scale $l$ is large when the membrane is weak compared to the rays and small when the membrane is strong. 
If $l$ is much longer than the fin length $L$, the influence of bending-stretching coupling on the fin stiffness can be ignored, but if $l$ is much shorter, the influence dominates. 
The coupling parameter $L/l$ characterizes the strength of this bending-stretching coupling and the membrane stiffness.
Furthermore, the fin stiffness depends on the purely geometric parameter $\gamma$, which characterizes the anisotropic bending of the individual rays.

The two asymptotic behaviours of membrane strength are understood in detail by considering the direction of ray bending.
When the membrane is weak and the membrane stretching force is negligible, the rays bend predominantly in the direction of the applied load, i.e.\ in the normal, soft direction. 
The fin stiffness in this limit, $R_0$, is simply the stiffness of an Euler-Bernoulli beam of length $L$ in that direction \cite{timoshenko1930strength},
\begin{align}
R_0 = \dfrac{3B_n}{L^3}.
\label{eqn:R_0}
\end{align}
At the other extreme when the membrane is strong, the membrane elongation is negligible, and the two rays bend predominantly perpendicular to the single membrane.
The fin stiffness in this limit, $R_\infty$, is 
\begin{align}
R_\infty = \dfrac{3\Bn}{L^3}\left( 1 + \gamma \right) = R_0 (1+\gamma). \label{eqn:Rinfty}
\end{align}
Depending on the anisotropy of the rays $\gamma$, the fin stiffness $R_\infty$ could be much larger than $R_0$.

To derive the complete dependence of the 2-ray fin stiffness on membrane elasticity, we solve for the membrane stretching.
Equation \eqref{eqn:2raystr} has four fundamental solutions in terms of the complex number $\za = e^{i\pi/4}$ and its complex conjugate $\zb$, namely $e^{\za x/l}$, $e^{\zb x/l}$, $e^{-\za x/l}$, and $e^{-\zb x/l}$.
Only two linear combinations $Z_{1,2}(x/l)$ of the four solutions satisfy the boundary conditions in equation \eqref{eqn:bc} at the fin base. 
In turn, the solution $\str(x/l)$ must be a linear combination of $Z_{1,2}(x/l)$ satisfying the conditions at the ray tips.
Algebraic manipulation of undetermined coefficients directly yield $\str(x)$, and the membrane stretching at $x=L$ would be given in terms of $\xi = L/l$, $Z_{1,2}(x/l)$, and their derivatives $X_{1,2}$ and $Y_{1,2}$ as 
\begin{subequations}
\begin{align}
& Z_1(\xi) =      \cosh \left(\zb \xi \right)   -       \cosh \left( \za \xi  \right),  \quad
&&Z_2(\xi) = \za  \sinh \left(\zb \xi \right)   -  \zb  \sinh \left( \za \xi  \right),\\
&X_1(\xi) =      Z_1''({\xi}),
&& X_2(\xi) =      Z_2''({\xi}), \quad 	\\
& Y_1(\xi) =      Z_1'''({\xi}) ,  \quad
&&Y_2(\xi) =      Z_2'''({\xi}). 
\end{align}
\end{subequations}
to obtain
\begin{align}
\str(L) = \dfrac{2f\sin\theta L^3}{\Bn} \eta(\xi), \quad 
\text{ where } \eta\left(\xi \right)   = \dfrac{3}{\xi^3}~\frac{X_2(\xi) Z_1(\xi) - X_1(\xi) Z_2(\xi)}{X_1(\xi) Y_2(\xi) - X_2(\xi) Y_1(\xi)}.
\label{eqn:eta}
\end{align}
The physical significance of $\eta(L/l)$ is apparent from equation \eqref{eqn:eta}; $\eta$ is the membrane stretching at the ray tips, normalized by the influence of the external point load $f$ acting on the membrane such that $\eta$ ranges from zero to unity.
The normal bending at the ray tips is given as
\begin{align}
w(L) =  \dfrac{fL^3}{3\Bn}\dfrac{\Bn\cos^2\theta + \eta\Bt\sin^2\theta}{\Bn\cos^2\theta+\Bt\sin^2\theta}, \quad
\label{eqn:R1prime} 
\end{align}
and the fin stiffness $R$ is given in terms of $\eta$ and $\gamma$ as
\begin{align}
R = \dfrac{f}{w(L)} =  \dfrac{3\Bn}{L^3}\dfrac{1+\gamma}{1+\gamma\eta}. \quad
\label{eqn:R1} 
\end{align}
To highlight modulation of the fin's stiffness between its two limits $R_0$ and $R_\infty$ as $L/l$ changes with curvature, we express the stiffness $R$ as an interpolation between the two limits, as 
\begin{align}
R =  (1-\alpha)R_0 + \alpha R_\infty, \quad
\label{eqn:R} 
\text{ where }\alpha = \dfrac{1-\eta}{1+\gamma\eta}.
\end{align}
Since $0<\eta<1$, $\alpha$ also ranges between zero and unity and behaves as an interpolation variable for stiffness.  
Figure \ref{fig:cs}(d) plots equation \eqref{eqn:R1} in terms of the coupling parameter $L/l$, whereas figure \ref{fig:cs}(e) plots equation \eqref{eqn:R} to demonstrate the influence of $\gamma$ on bending-stretching coupling. 

The dependence of fin stiffness on membrane stiffness $L/l$ and ray anisotropy $\gamma$ are understood as follows, and demonstrated in figures~\ref{fig:cs}(d) and (e).
Consider first the case of nearly isotropic rays ($\gamma \approx 1$, blue curves in figures~\ref{fig:cs}(d) and (e)).
As the membrane is slightly stiffened but still weak ($L/l \ll 1$), the fin stiffness begins to increase over $R_0$, and is found from \eqref{eqn:R} and \eqref{eqn:eta} as $R \approx R_0 (1 + 37 k L^4 \sin^2\theta /210 B_n + \dots)$.
In this regime, the fin stiffness is determined by a combination of the normal bending and membrane stretching, but the weak membrane does not engage the tangential bending mode of the rays, and fin stiffness is independent of the tangential bending rigidity.
As the membrane stiffness increases so much that $(L/l \gg 1)$, the fin stiffness approaches $R_\infty$ approximately as $R \approx R_\infty ( 1 - 3\sqrt{2} \gamma (l/L)^3 + \dots)$.
Here, the rays bend so that the membrane does not stretch, even if that engages the rays' stiffer tangential bending mode.

A different picture emerges for highly anisotropic rays ($\gamma \gg 1$, see curves for $\gamma$=10, 100 and 1000 in figures~\ref{fig:cs}(d) and (e)).
For very weak membranes ($L/l \ll 1$), the stiffness remains independent of tangential bending and is given by $R \approx R_0 (1 + 37 k L^4 \sin^2\theta /210 B_n + \dots)$ as before.
However, because these rays are much stiffer to tangential bending, the membrane force also results in little tangential bending in the intermediate range of membrane stiffnesses corresponding to $L/l \gg 1$ but  $ \eta \gamma \ll 1$.
Using the asymptotic form of $\eta(\xi) \propto \xi^{-3}$ for $\xi \gg 1$, this criteria may be translated to $1 \ll L/l \ll \gamma^{1/3}$.
In this intermediate range, the fin stiffness may be approximated as $R/R_0 \approx (L/l)^3/3\sqrt{2}$ and gains a dependence on the anisotropy parameter.
The stiffness in this range remains smaller than $R_\infty$.
As the membrane stiffness further increases, a stage is reached where the membrane force is strong enough to bend the rays in the tangential direction.
It is in this regime that the tangential bending of rays appreciably reduces the stretching of the membrane, the interpolation parameter $\alpha$ transitions between zero and unity, and the fin stiffness is a significant fraction of $R_\infty$.
Based on \eqref{eqn:R}, this implies $\gamma \eta(L/l) = O(1)$, i.e.\ $L/l = O(\gamma^{1/3})$. 
Finally, for an even larger membrane stiffness, corresponding to the regime where $L/l \gg \gamma^{1/3}$, $\alpha$ approaches unity, and similar to the case of isotropic rays, the stiffness approaches $R_\infty$ approximately as $R \approx R_\infty ( 1 - 3\sqrt{2} \gamma (l/L)^3 + \dots )$.

Whether the fin is highly anisotropic or nearly isotropic, the criteria for spanning the range of fin stiffness compares $L/l$ to $\gamma^{1/3}$.
The important distinction between isotropic and anisotropic rays is that the anisotropic bending of the rays leads to a much stiffer fin, i.e. $R_\infty/R_0 = 1+\gamma \gg 1$.

\subsection{Design criterion for stiffness modulation}
The range of stiffness accessible to a fin depends on the range of curvature, from zero to $\theta_\text{max}$, it can reasonably impose on itself.
Changing curvatures simultaneously changes the coupling parameter $L/l$ and the ray anisotropy $\gamma$.
Here we derive the influence of the curvature on these parameters and apply the result of the previous subsection to determine the range of stiffness accessible to a fin.
If $\left( \frac{L}{l} \right)_\text{max}$ is much less than $\left( \gamma_\text{max} \right)^{1/3}$, where `max' denotes the respective values at $\theta_\text{max}$, then the fin can only access the minimal stiffness $R_0$ determined by the normal stiffness of rays, and stiffening by curvature would be impractical in such a fin.
If $\left( \frac{L}{l} \right)_\text{max} $ is comparable to or greater than $\left( \gamma_\text{max} \right)^{1/3}$, the fin can span at least an appreciable fraction of the range between the minimal stiffness $R_0$ and maximal stiffness $R_\infty$,
and this condition is the design criterion by which to check if stiffening by curvature is practical.
Since $R_\infty$ also depends on curvature, we limit the rest of this analysis to $\gamma_\text{max} \gg 1$ in the interest of the case that $R_\infty \gg R_0$.
We show in the appendix that choosing $\theta_\text{max}=\pi/4$ for the 2-ray fin provides a quantitative comparison with the parameters for the $N$-ray fin, in which case
\begin{align}
 \left( \dfrac{L}{l} \right)_\text{max} = L \left( \dfrac{k}{\Bn} + \dfrac{k}{\Bt} \right)^{1/4} \qquad \text{ and } \qquad \gamma_\text{max} = \dfrac{\Bt}{\Bn}.
 \label{eqn:Loverlmax}
\end{align}




The design criterion for stiffening compares the maximum coupling a fin can achieve $\left( \frac{L}{l} \right)_\text{max}$ to $\left( \gamma_\text{max} \right)^{1/3}$.
If $\left( \frac{L}{l} \right)_\text{max}$ is comparable to $\left( \gamma_\text{max} \right)^{1/3}$, then the fin may access the transition regime between $R_0$ and $R_\infty$ and span an appreciable fraction of the range. 
A dash black line in figure \ref{fig:cs}(e) is representative of $\left( \frac{L}{l} \right)_\text{max} = \left( \gamma_\text{max} \right)^{1/3}$.
On the other hand, if $\left( \frac{L}{l} \right)_\text{max} \gg \left( \gamma_\text{max} \right)^{1/3}$, the fin is past the transition and accesses a stiffness marginally close to $R_\infty$.
It can practically span the entire range of stiffness.
Under the assumption that $\Bt \gg \Bn$ (or $\gamma \gg 1$), these conditions simplify to
\begin{align}
\begin{split}
 &L \left(\dfrac{k}{\Bn}\right)^{1/4} \gg \left(\dfrac{\Bt}{\Bn}\right)^{1/3} \qquad \text{ for spanning the whole range and} \\
 &L \left(\dfrac{k}{\Bn}\right)^{1/4} \sim \left(\dfrac{\Bt}{\Bn}\right)^{1/3} \qquad \text{ for spanning an appreciable fraction of the range},
\end{split}
\end{align}
where $\sim$ stands for ``approximately in the range of''. Analogous analysis for an $N$-ray fin is presented in the Appendix A, which yields the result
\begin{align}
\begin{split}
 &L \left(\dfrac{\pi^2 k}{N^2 \Bn}\right)^{1/4} \gg \left(\dfrac{\Bt}{\Bn}\right)^{1/3} \qquad \text{ for spanning the whole range and} \\
 &L \left(\dfrac{\pi^2 k}{N^2 \Bn}\right)^{1/4} \sim \left(\dfrac{\Bt}{\Bn}\right)^{1/3} \qquad \text{ for spanning an appreciable fraction of the range},
\end{split}
 \label{eqn:findesigncriteria}
\end{align}
where the extra factor of $(\pi/N)^{1/2}$ accounts for the effect of distributing the bending-stretching coupling over $N$ rays.
We illustrate the application of this criterion to a Mackerel pectoral fin in \S\ref{sec:verification}, and propose that the morphology of the fin could permit the Mackerel to modulate its stiffness by at least 300\%.

\subsection{Morphological and functional curvature}
The principle that underlies curvature-induced stiffening is the kinematic coupling of the bending of rays and the stretching of membranes. 
In addition to the morphological curvature considered thus far, this kinematic coupling can also be accomplished in a flat fin by systematic misalignments of the soft bending directions of adjacent rays. 
We call this ``functional curvature''.
It is evident in the pectoral fin of a Mackerel as seen in figure \ref{fig:Springs}(d-e), which shows $\mu$CT cross-sections of a fin at 20\% - 30\% along its length from the base while being held flat between two plates (figure \ref{fig:Springs}(c)). 
The effective curvature of rayed fins is a mixture of both forms of curvature, and information of both the overall shape of the fin and orientation of rays is necessary to adequately model fin stiffness.

We present here the qualitative similarities between the two types of curvatures, although the quantitative and specific dependences necessarily differ.
Using the first (leftmost) membrane shown in figure \ref{fig:Springs}(a) as an example, the geometric relation in the case of morphological curvature is
\begin{align}
&\str_1 = [w_1 \sin\theta_1 + w_2 \sin\theta_1] + [v_1 \cos\theta_1 - v_2\cos\theta_1],
\end{align}
where the elongation of the first membrane is $\str_1$, the normal displacements of the first and second ray are $w_1$ and $w_2$ respectively, the tangential displacements are $v_1$ and $v_2$, and the misalignment of membrane 1 and its neighbouring rays is $\theta_1$. 
Analogously, the first membrane of figure \ref{fig:Springs}(b) in the case of functional curvature has a geometric relation as
\begin{align}
&\str_1 = [w_1 \sin\phi_1 - w_2 \sin\phi_2] + [v_1 \cos\phi_1 - v_2\cos\phi_2],
\end{align}
where $\phi_1$ and $\phi_2$ are the angles made by the first membrane with the first and second rays respectively. 
In both curvatures, membrane stretching is a combination of displacements of the neighboring rays.
The linear weights differ, and therefore the mathematical solution may need to be obtained numerically, but the fin is expected to stiffen due to bending-stretching coupling.


\section{Comparison with sample fish fins}
\label{sec:verification}
\begin{figure}[t]
\centerline{\includegraphics[width=100mm]{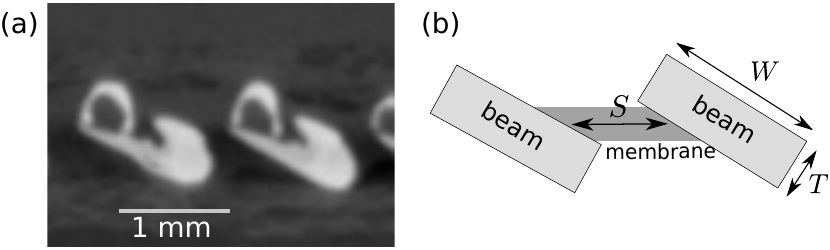}}
\caption{Schematic aiding estimation of various parameters of the simple caricature of fish fin cross-section. 
(a) A representative, composite image of rays 9 and 10 from figure \ref{fig:Springs}(d).
(b) A schematic of the simple caricature approximating the rays to have a rectangular cross-section and the membrane to be a parallelogram.}
\label{fig:FishFinDesign}
\end{figure}

We compare the design criteria derived in \eqref{eqn:findesigncriteria} with geometric measurements obtained from cross-sections of a Mackerel pectoral fin shown in figure \ref{fig:FishFinDesign}.
For this comparison, we treat the ray cross-sections as rectangles of dimensions $W \times T$, and bending rigidities
\begin{align}
 \Bn \sim \dfrac{E_b T^3 W}{12 (1-\nu^2)} \qquad \text{ and } \qquad \Bt \sim \dfrac{E_b T W^3} {12 (1-\nu^2)},
 \label{eqn:abc}
\end{align}
where $E_b$ is the elastic modulus of bone that constitutes the rays, and $\nu$ is its Poisson ratio.
The dependence on the Poisson ratio on the subsequent analysis is weak, and we henceforth neglect it.
Therefore, the ratio $\gamma = \Bt/\Bn \sim W^2/T^2$ quantifies the bound on the anisotropy parameter. 
Similarly, the membrane cross-section is modelled to be a parallelogram, which spaces the rays an amount $S$, leading to an estimate for its elasticity constant 
\begin{align}
k \sim E_c \dfrac{T}{S}, 
\end{align}
where $E_c$ is the elastic modulus of the material (collagen) that makes up the membrane.
Using these expressions, the two sides of the criteria \eqref{eqn:findesigncriteria} for whether the fin design allows it to span the range of stiffnesses from $R_0$ to $R_\infty$ may be written as
\begin{align}
\text{l.h.s.} = L \left( \dfrac{\pi^2 k}{N^2 \Bn} \right)^{1/4} \approx L \left( \dfrac{12 \pi^2 E_c}{N^2 E_b T^2 W S } \right)^{1/4} \qquad \text{ and } \qquad \text{r.h.s.} = \left(\dfrac{\Bt}{\Bn} \right)^{1/3} \approx \left( \dfrac{W}{T} \right)^{2/3},
\label{eqn:geomdesigncriteria}
\end{align}
where l.h.s. and r.h.s. stands for the left and right hand side respectively.

Using measurements for the geometric parameters of sampled rays in the Mackerel pectoral fin as shown in figure \ref{fig:Springs}(c), and representative values for the elastic moduli in equation \eqref{eqn:geomdesigncriteria}, we deduce the importance of bending-stretching coupling for the fish fin.
Translation of sampled cross-sections to the simple model of a rectangular ray as shown in figure \ref{fig:FishFinDesign}, yields $W = 0.8-1.2$ mm, $T = 0.5-0.6$ mm, and $S =0.2-0.6$ mm. 
We use $L=9$ mm for the fin's length and $N \approx 10$ for the number of rays in the fin.
Furthermore, $E_c$ lies in the range 3-12 GPa \cite{wenger2007mechanical}, while $E_b$ lies in the range 4-20 GPa \cite{reilly1974elastic, choi1990elastic, rho1998mechanical}.
Based on these estimates, we provide a range for both sides of the design criteria shown in equation \eqref{eqn:geomdesigncriteria}.
The enhancement of stiffness possible through the bending-stretching coupling is by a factor $1+\gamma = 1 + W^2/T^2$, realized if the criteria in equation \eqref{eqn:geomdesigncriteria} are satisfied.  
Substituting these numbers in \eqref{eqn:geomdesigncriteria} yields that the l.h.s. lies in the range of 10-26, while r.h.s. lies in the range of 1.6-3.5.
The l.h.s. is more than the r.h.s. by almost a factor of 3 at worst and an order of magnitude larger at best.
Therefore, bending-stretching coupling is influential at the cross-sections where these estimates were made.

Based on these observations, we conclude that the bending-stretching coupling induced by curvature influences the Mackerel pectoral fin stiffness.
This fin may be capable of modulating its stiffness by a factor of $1+W^2/T^2$ ranging from 300--700\% relative to the normal stiffness of its rays. 
Similar analysis performed with other fins may be used to determine whether bending-stretching coupling is relevant for those fins, and whether the implications of this article apply to them.

\section{Discussion}
Our work adds a critical piece to the existing knowledge of fish fin structure and function.
Each fin ray is known to be composed of a bi-layer structure, and differentially pulling the base of these layers curls the ray \cite{geerlink1986relation,alben2007mechanics}.
It has been assumed that the rays act as girders \cite{mccutchen1970trout} and are the sole structural elements that reinforce the fin. 
The surface formed by an assembly of curled rays determines the shape of the fin surface and the bending of rays is considered as the primary mechanism imparting stiffness to the fin \cite{tangorra2010effect}.
The membrane interconnecting the rays is tacitly assumed to contribute negligibly to the bending stiffness \cite{zhu2008propulsion,shoele2009fluid,shoele2010numerical}.
Measurement of fin stiffness has therefore been considered synonymous with the measurement of bending rigidity of the rays themselves \cite{tangorra2010effect}, and no other parameters are considered.
Our analysis points to the bending-stretching coupling as an important element for structurally reinforcing the fin.
The definition of fin curvature, and its contribution to fin stiffness through a composite ray-membrane structure are the main results of this article.

Observations from $\mu$CT imaging of the Mackerel fin indicate the presence of a functional curvature.
Even when the pectoral fin is held flat, there remain misalignments of principal bending direction of neighbouring rays built into the fin's skeletal structure.
The misalignments are observed in the cross-section of rays taken normal to each ray's axis at varying fractions ($20\% - 30\%$) of the total fin length (figure \ref{fig:Springs}(d-e)) in a Mackerel pectoral fin. 
These internal misalignments functionally mimic an external transverse curvature and characterize a functional curvature, i.e.\ morphological elements that couple bending and stretching.

The basis of our results lies in principles of mechanics and mathematics, and we chose the Mackerel pectoral fin to illustrate these concepts owing to its ease of availability.
Are the results we find applicable across fish species?
A comparative study of fin skeletal cross-sections from diverse fish species for anisotropy of bending moduli and misalignments of principal axes in adjoining rays is needed.
The design criteria derived in \S\ref{sec:verification} are useful to determine whether transverse curvature could stiffen the fin.
Curvature without a sufficiently stiff membrane does not stiffen the fin, and the design criteria define this relationship.
If a functional curvature is present, then it must also be included to determine the fin stiffness, and incorporated in the mathematical models of fin fluid-structure interactions to adequately model the propulsive abilities of rayed-fin fishes.
The mechanical nature of the argument, the geometric origin of the coupling between bending of rays to stretching of membrane, and the shared genetic and developmental processes in ray-finned fish allows us to generalize our results to other fish-fin structure with rays and membranes. 

Quantitative deviations of fish fins from our rayed-fin models introduce complications but our analysis remains qualitatively applicable for real fins.
Individual rays tend to segment, taper and branch towards their distal ends \cite{flammang2013functional}.
Both the segmented and branched regions of rays introduce areas of non-bony tissue that change the material properties of rays, while tapering changes the ray's width and thickness.
These cause variations in local parameters, such as the bending moments, membrane thicknesses, and the spacing between rays.
As a result, the strength of bending-stretching-coupling also varies. 
In our estimates of the design criterion in equations (\ref{eqn:abc} - \ref{eqn:geomdesigncriteria}), we sampled non-segmented and non-branched regions in order to approximate the ray as bony material.
Checking for the design criterion, and the strength of bending-stretching coupling in the segmented or branched regions would require a characterization of the correct bending rigidities. 
The principle of curvature-induced bending-stretching coupling, however, remains equally applicable.

Curvature-induced stiffening provides an alternative explanation for the enhancement of thrust due to ``cupping'' of biomimetic fins. 
Cupping is the term for kinematics of the fin that transversely curves into the flow as it flaps \cite{lauder2002forces,tytell2006median,flammang2008speed,esposito2012robotic}.
This shape is achieved by oscillating the first and the last fin rays with greater amplitude than the rest, and/or with a phase that leads the one in between.
The fluid vorticity pattern observed in the wake of cupped fins bears the signature of enhanced thrust.
We postulate that the enhanced thrust is a result of stiffening the fin to out-of-plane bending deformation.
The force transmitted by the fin depends on the acceleration of the surrounding fluid it can produce; greater the fluid acceleration, greater the transmitted force.
A soft fin, when actuated at its base, would deform under the transmitted load instead of pushing the surrounding fluid.
Thus flexibility of the fin limits its ability to transmit forces.
In contrast, a stiffer fin is able to transmit stronger forces before deforming; independent calculations and experiments varying the flexibility of the oscillating fins confirm the increase in maximum thrust generated by stiffer fins \cite{zhu2008propulsion,alben2008optimal,michelin2009resonance,shoele2009fluid,spagnolie2010surprising,tangorra2010effect,park2012kinematic,esposito2012robotic,moore2015torsional,masoud2010resonance}.
It may be for this reason that cupping is commonly observed in fish caudal fin kinematics \cite{bainbridge1963caudal,geerlink1986relation,lauder2000function}.

The rayed-fin analysis identifies key material parameters and structural elements that are critical for accurately predicting propulsive performance, and should be incorporated to improve mathematical models. 
Most models of propulsion simply ignore the three-dimensional nature of the fluid-structure interaction \cite{alben2008optimal,michelin2009resonance,spagnolie2010surprising,shoele2012leading}.
Those that do incorporate three-dimensionality \cite{zhu2008propulsion,shoele2009fluid,shoele2010numerical} treat the rays as non-linear Euler-Bernoulli beams made of linear elastic material but do not consider anisotropic bending moduli. 
Furthermore, the membrane is assumed to be inextensible in the transverse direction.
Part of the reason for these simplifications may be the lack of experimental characterization of the anisotropic nature of ray bending or of the elastic properties of the membrane.
By ignoring the anisotropy, the stretching of the membrane, the attachment of the membrane to the rays, and the resulting torsion, bending-stretching coupling induced by curvature is inadequately modelled.
As our analysis shows, these assumptions imply operation in a very limited space of the bending-stretching coupling. 
For this reason, the elastic properties of the interconnecting membrane, the anisotropic bending moduli and the torsional stiffness of the rays are important to experimentally quantify in fish and incorporate in a mathematical model before their predictions are representative.

Another important factor that needs to be better understood is how an external transverse curvature arises in a fish fin.
The musculature at the base of the fin could actively deform the fin in a curved shape.
The fish may manipulate the joints at the base of the fin that curves it in the same manner as we curve a sheet of paper holding it only at one edge.
The action, for example, of the interradialis muscles distributed at the base of caudal fins in concert with the basal skeletal structure \cite{flammang2008speed,flammang2009caudal} may bring about such an effect.
Yet another manner of inducing curvature could be through non-uniform bending of fin rays via differential actuation of their hemitrichs \cite{geerlink1986relation,alben2007mechanics}. 
The shape of the fin is then determined by the combined influence of bilamellar ray bending, membrane stretching, and ray torsion. 
The specific manner of attachment between the membrane and the adjacent rays may cause the rays to twist along their axes, thereby complicating the calculation of bending force on account of the anisotropic bending moduli. 
Such torsion of the rays is therefore an important effect to evaluate and include in computational models which omit its effect \cite{zhu2008propulsion,shoele2009fluid,shoele2010numerical,shoele2010numerical}.

The rayed-fin analysis forms the basis of a preliminary mechanical design of biomimetic fins to span a specified stiffness range.
The rays are first designed by selecting a cross-section so that the minimum stiffness, $R_0$ from \eqref{eqn:R_0}, and the maximum stiffness, $R_\infty$ from \eqref{eqn:Rinfty}, take on desired values. 
The membrane stiffness is then determined by selecting $S$ so that the criterion \eqref{eqn:geomdesigncriteria} implies that the whole range of the designed stiffness is spanned.

The analysis also forms a mechanical basis for the hypothesis that underlying the anatomy of the fish fin structure and its ability to modulate stiffness \cite{videler1977mechanical} is the need to be manoeuvrable. 
Such manoeuvrability is reflected in a variety of propulsive modes \cite{webb1994biology,sfakiotakis1999review} the fish use, or in their ability to rapidly deform their body with or without burst of momentum exchange with the surrounding fluid \cite{epps2007impulse}. 
Fish can also benefit from flow structures in the environment by either reducing their energy consumption or increasing their swimming speed \cite{liao2007review, webb2010turbulence, mclaughlin1998going, hinch1998swim, hinch2000optimal}.
These abilities rely upon the possibility of selectively engaging or disengaging the mechanical force transmission between the body and the surrounding fluid.
The curvature-induced stiffness modulation is the mechanical design that imparts the fins with this ability.

At the core of the mechanical design is the principle that a surface with small mechanical impedance moving relative to surrounding fluid experiences a small hydrodynamic force.
A surface that easily bends and deforms can only support a weak external force.
This principle is manifest in theoretical, computational, and experimental studies of fluid-structure interaction as a limit on the hydrodynamic force generated by flexible propulsive surfaces \cite{zhu2008propulsion,alben2008optimal,michelin2009resonance,shoele2009fluid,spagnolie2010surprising,tangorra2010effect,park2012kinematic,esposito2012robotic,moore2015torsional}.
Thus, the force transmission between the fish body and the fluid is disengaged when the fin is flat and soft, and engaged when the fin is curved and stiff.
Based on the above argument, it is likely that the range of stiffnesses spanned by fins of a species is selected based on its preferred behaviour and correlates with the ecological niche it occupies.

Inducing a curvature by purely basal actuation is a simple solution to the problem of stiffness modulation in propulsive surfaces. 
The alternative strategy of incorporating muscular actuation throughout the propulsive surface and using neural control to manipulate stiffness is also possible, but appears to suffer many drawbacks compared to the curvature-induced stiffening.
The additional mass of the actuators imposes a lower bound on the impedance of such structures.
This issue is further exacerbated by the added muscle needed to move the hypothetical muscles within the propulsive surface themselves!
The logistical problem of signal and energy transmission networks (e.g.\ innervation and vasculature) increases the complexity and fragility of the design.
Based on these considerations, the curvature-induced stiffening surface appears to have performance advantages, which could form the basis of a bio-inspired propulsive appendage.

\section{Conclusion}
We have presented an essential model analysing the stiffening of a fish fin when curved transverse to the fin rays.
The stiffening occurs because the curvature couples the out-of-plane bending of rays to the in-plane stretching of the elastic membrane connecting the rays.
As a function of its geometric and material parameters, we have calculated the range of stiffnesses spanned by a fin with changing curvature.
We postulate that this range evolves in response to the selection pressures imposed by the ecological niche the fish species occupies.
Furthermore, our analysis of such stiffening has led to a functional definition of curvature itself, which we have found to be present even when the fin appears geometrically flat.
We have made many simplifying assumptions in our analysis to derive insight into the mechanics of fish fin stiffness. 
There undoubtedly are quantitative deviation from our predictions due to violations of our assumptions by real fish fin structures. 
Furthermore, there is variability in the material properties of tissues in the fin.
Yet, our mathematical model captures the principle that transcends these complications and uncertainties; the misaligned directions of the anisotropic bending modulus engages the stretching of the intervening membrane. 

\section*{Data Accessibility}
The X-ray image stack for the Mackerel fin is available from the Dryad Data Repository \cite{nguyen2017curvature:data}

\section*{Competing interests}
We have no competing interests.

\section*{Author contributions statement}

S.M.,\ M.V.\ and M.M.B.\ conceived the analysis, K.N.,\ N.Y.,\ and S.M.\ conducted the analysis, and all authors wrote and reviewed the manuscript.

\section*{Acknowledgements}
We acknowledge help from Biodiversity and Biocomplexity Unit, OIST directed by Prof. Evan Economo for sharing the equipment for Micro CT measurements.

\section*{Funding}
This research received funding from the Human Frontier Science Program.

\bibliographystyle{abbrvnamed}
\bibliography{refs-rspi}

\begin{thebibliography}{}

\bibitem[\protect\citeauthoryear{Alben \bgroup \em et al.\egroup
  }{2007}]{alben2007mechanics}
S.~Alben, P.~G. Madden, and G.~V. Lauder.
\newblock {The mechanics of active fin-shape control in ray-finned fishes}.
\newblock {\em Journal of The Royal Society Interface}, 4(13):243--256, 2007.

\bibitem[\protect\citeauthoryear{Alben}{2008}]{alben2008optimal}
S.~Alben.
\newblock {Optimal flexibility of a flapping appendage in an inviscid fluid}.
\newblock {\em Journal of Fluid Mechanics}, 614:355--380, 2008.

\bibitem[\protect\citeauthoryear{Bainbridge}{1963}]{bainbridge1963caudal}
R.~Bainbridge.
\newblock {Caudal fin and body movement in the propulsion of some fish}.
\newblock {\em Journal of Experimental Biology}, 40(1):23--56, 1963.

\bibitem[\protect\citeauthoryear{Choi \bgroup \em et al.\egroup
  }{1990}]{choi1990elastic}
K.~Choi, J.~Kuhn, M.~Ciarelli, and S.~Goldstein.
\newblock The elastic moduli of human subchondral, trabecular, and cortical
  bone tissue and the size-dependency of cortical bone modulus.
\newblock {\em Journal of Biomechanics}, 23(11):1103--1113, 1990.

\bibitem[\protect\citeauthoryear{Epps and Techet}{2007}]{epps2007impulse}
B.~P. Epps and A.~H. Techet.
\newblock {Impulse generated during unsteady maneuvering of swimming fish}.
\newblock {\em Experiments in Fluids}, 43(5):691--700, 2007.

\bibitem[\protect\citeauthoryear{Esposito \bgroup \em et al.\egroup
  }{2012}]{esposito2012robotic}
C.~J. Esposito, J.~L. Tangorra, B.~E. Flammang, and G.~V. Lauder.
\newblock {A robotic fish caudal fin: effects of stiffness and motor program on
  locomotor performance}.
\newblock {\em The Journal of Experimental Biology}, 215(1):56--67, 2012.

\bibitem[\protect\citeauthoryear{Flammang and Lauder}{2008}]{flammang2008speed}
B.~E. Flammang and G.~V. Lauder.
\newblock {Speed-dependent intrinsic caudal fin muscle recruitment during
  steady swimming in bluegill sunfish, Lepomis macrochirus}.
\newblock {\em Journal of Experimental Biology}, 211(4):587--598, 2008.

\bibitem[\protect\citeauthoryear{Flammang and
  Lauder}{2009}]{flammang2009caudal}
B.~Flammang and G.~Lauder.
\newblock {Caudal fin shape modulation and control during acceleration, braking
  and backing maneuvers in bluegill sunfish, Lepomis macrochirus}.
\newblock {\em Journal of Experimental Biology}, 212(2):277--286, 2009.

\bibitem[\protect\citeauthoryear{Flammang \bgroup \em et al.\egroup
  }{2013}]{flammang2013functional}
B.~E. Flammang, S.~Alben, P.~G. Madden, and G.~V. Lauder.
\newblock {Functional morphology of the fin rays of teleost fishes}.
\newblock {\em Journal of Morphology}, 274(9):1044--1059, 2013.

\bibitem[\protect\citeauthoryear{Geerlink and
  Videler}{1986}]{geerlink1986relation}
P.~Geerlink and J.~Videler.
\newblock The relation between structure and bending properties of teleost fin
  rays.
\newblock {\em Netherlands Journal of Zoology}, 37(1):59--80, 1986.

\bibitem[\protect\citeauthoryear{Gosline}{1971}]{gosline1971functional}
W.~A. Gosline.
\newblock {\em {Functional morphology and classification of teleostean
  fishes}}.
\newblock University Press of Hawaii, 1971.

\bibitem[\protect\citeauthoryear{Hinch and Rand}{1998}]{hinch1998swim}
S.~G. Hinch and P.~S. Rand.
\newblock {Swim speeds and energy use of upriver-migrating sockeye salmon
  (Oncorhynchus nerka): role of local environment and fish characteristics}.
\newblock {\em Canadian Journal of Fisheries and Aquatic Sciences},
  55(8):1821--1831, 1998.

\bibitem[\protect\citeauthoryear{Hinch and Rand}{2000}]{hinch2000optimal}
S.~G. Hinch and P.~S. Rand.
\newblock {Optimal swimming speeds and forward-assisted propulsion:
  energy-conserving behaviours of upriver-migrating adult salmon}.
\newblock {\em Canadian Journal of Fisheries and Aquatic Sciences},
  57(12):2470--2478, 2000.

\bibitem[\protect\citeauthoryear{Lauder and Drucker}{2002}]{lauder2002forces}
G.~V. Lauder and E.~G. Drucker.
\newblock {Forces, fishes, and fluids: hydrodynamic mechanisms of aquatic
  locomotion}.
\newblock {\em Physiology}, 17(6):235--240, 2002.

\bibitem[\protect\citeauthoryear{Lauder and
  Drucker}{2004}]{lauder2004morphology}
G.~V. Lauder and E.~G. Drucker.
\newblock {Morphology and experimental hydrodynamics of fish fin control
  surfaces}.
\newblock {\em IEEE Journal of Oceanic Engineering}, 29(3):556--571, 2004.

\bibitem[\protect\citeauthoryear{Lauder}{1989}]{lauder1989caudal}
G.~V. Lauder.
\newblock {Caudal fin locomotion in ray-finned fishes: historical and
  functional analyses}.
\newblock {\em American Zoologist}, 29(1):85--102, 1989.

\bibitem[\protect\citeauthoryear{Lauder}{2000}]{lauder2000function}
G.~V. Lauder.
\newblock {Function of the caudal fin during locomotion in fishes: kinematics,
  flow visualization, and evolutionary patterns}.
\newblock {\em American Zoologist}, 40(1):101--122, 2000.

\bibitem[\protect\citeauthoryear{Liao}{2007}]{liao2007review}
J.~C. Liao.
\newblock {A review of fish swimming mechanics and behaviour in altered flows}.
\newblock {\em Philosophical Transactions of the Royal Society of London B:
  Biological Sciences}, 362(1487):1973--1993, 2007.

\bibitem[\protect\citeauthoryear{Lund}{1967}]{lund1967analysis}
R.~Lund.
\newblock {\em {An analysis of the propulsive mechanisms of fishes, with
  reference to some fossil actinopterygians}}.
\newblock Carnegie Museum, 1967.

\bibitem[\protect\citeauthoryear{Masoud and
  Alexeev}{2010}]{masoud2010resonance}
H.~Masoud and A.~Alexeev.
\newblock Resonance of flexible flapping wings at low reynolds number.
\newblock {\em Physical Review E}, 81(5):056304, 2010.

\bibitem[\protect\citeauthoryear{McCutchen}{1970}]{mccutchen1970trout}
C.~McCutchen.
\newblock {The trout tail fin: a self-cambering hydrofoil}.
\newblock {\em Journal of Biomechanics}, 3(3):271--281, 1970.

\bibitem[\protect\citeauthoryear{McLaughlin and
  Noakes}{1998}]{mclaughlin1998going}
R.~L. McLaughlin and D.~L. Noakes.
\newblock {Going against the flow: an examination of the propulsive movements
  made by young brook trout in streams}.
\newblock {\em Canadian Journal of Fisheries and Aquatic Sciences},
  55(4):853--860, 1998.

\bibitem[\protect\citeauthoryear{Michelin and
  {Llewellyn-Smith}}{2009}]{michelin2009resonance}
S.~Michelin and S.~G. {Llewellyn-Smith}.
\newblock {Resonance and propulsion performance of a heaving flexible wing}.
\newblock {\em Physics of Fluids}, 21(7):071902, 2009.

\bibitem[\protect\citeauthoryear{Moore}{2015}]{moore2015torsional}
M.~N.~J. Moore.
\newblock Torsional spring is the optimal flexibility arrangement for thrust
  production of a flapping wing.
\newblock {\em Physics of Fluids}, 27(9):091701, 2015.

\bibitem[\protect\citeauthoryear{Nguyen \bgroup \em et al.\egroup
  }{2017}]{nguyen2017curvature:data}
K.~Nguyen, N.~Yu, M.~Bandi, M.~Venkadesan, and S.~Mandre.
\newblock {\em Data from: Curvature-induced stiffness of fish fin}.
\newblock Dryad Digital Repository, 2017.

\bibitem[\protect\citeauthoryear{Park \bgroup \em et al.\egroup
  }{2012}]{park2012kinematic}
Y.-J. Park, U.~Jeong, J.~Lee, S.-R. Kwon, H.-Y. Kim, and K.-J. Cho.
\newblock {Kinematic condition for maximizing the thrust of a robotic fish
  using a compliant caudal fin}.
\newblock {\em IEEE Transactions on Robotics}, 28(6):1216--1227, 2012.

\bibitem[\protect\citeauthoryear{Reilly \bgroup \em et al.\egroup
  }{1974}]{reilly1974elastic}
D.~T. Reilly, A.~H. Burstein, and V.~H. Frankel.
\newblock The elastic modulus for bone.
\newblock {\em Journal of Biomechanics}, 7(3):271--275, 1974.

\bibitem[\protect\citeauthoryear{Rho \bgroup \em et al.\egroup
  }{1998}]{rho1998mechanical}
J.-Y. Rho, L.~Kuhn-Spearing, and P.~Zioupos.
\newblock Mechanical properties and the hierarchical structure of bone.
\newblock {\em Medical Engineering \& Physics}, 20(2):92--102, 1998.

\bibitem[\protect\citeauthoryear{Rosic \bgroup \em et al.\egroup
  }{2017}]{rosic2017performance}
M.-L.~N. Rosic, P.~J. Thornycroft, K.~L. Feilich, K.~N. Lucas, and G.~V.
  Lauder.
\newblock Performance variation due to stiffness in a tuna-inspired flexible
  foil model.
\newblock {\em Bioinspiration \& Biomimetics}, 12(1):016011, 2017.

\bibitem[\protect\citeauthoryear{Sfakiotakis \bgroup \em et al.\egroup
  }{1999}]{sfakiotakis1999review}
M.~Sfakiotakis, D.~M. Lane, and J.~B.~C. Davies.
\newblock Review of fish swimming modes for aquatic locomotion.
\newblock {\em IEEE Journal of Oceanic Engineering}, 24(2):237--252, 1999.

\bibitem[\protect\citeauthoryear{Shoele and Zhu}{2009}]{shoele2009fluid}
K.~Shoele and Q.~Zhu.
\newblock {Fluid--structure interactions of skeleton-reinforced fins:
  performance analysis of a paired fin in lift-based propulsion}.
\newblock {\em Journal of Experimental Biology}, 212(16):2679--2690, 2009.

\bibitem[\protect\citeauthoryear{Shoele and Zhu}{2010}]{shoele2010numerical}
K.~Shoele and Q.~Zhu.
\newblock {Numerical simulation of a pectoral fin during labriform swimming}.
\newblock {\em Journal of Experimental Biology}, 213(12):2038--2047, 2010.

\bibitem[\protect\citeauthoryear{Shoele and Zhu}{2012}]{shoele2012leading}
K.~Shoele and Q.~Zhu.
\newblock {Leading edge strengthening and the propulsion performance of
  flexible ray fins}.
\newblock {\em Journal of Fluid Mechanics}, 693:402--432, 2012.

\bibitem[\protect\citeauthoryear{Spagnolie \bgroup \em et al.\egroup
  }{2010}]{spagnolie2010surprising}
S.~E. Spagnolie, L.~Moret, M.~J. Shelley, and J.~Zhang.
\newblock {Surprising behaviors in flapping locomotion with passive pitching}.
\newblock {\em Physics of Fluids}, 22(4):041903, 2010.

\bibitem[\protect\citeauthoryear{Tangorra \bgroup \em et al.\egroup
  }{2010}]{tangorra2010effect}
J.~L. Tangorra, G.~V. Lauder, I.~W. Hunter, R.~Mittal, P.~G. Madden, and
  M.~Bozkurttas.
\newblock {The effect of fin ray flexural rigidity on the propulsive forces
  generated by a biorobotic fish pectoral fin}.
\newblock {\em Journal of Experimental Biology}, 213(23):4043--4054, 2010.

\bibitem[\protect\citeauthoryear{Timoshenko}{1930}]{timoshenko1930strength}
S.~Timoshenko.
\newblock {\em {Strength of materials}}.
\newblock New York, 1930.

\bibitem[\protect\citeauthoryear{Tytell}{2006}]{tytell2006median}
E.~D. Tytell.
\newblock {Median fin function in bluegill sunfish Lepomis macrochirus:
  streamwise vortex structure during steady swimming}.
\newblock {\em The Journal of Experimental Biology}, 209(8):1516--1534, 2006.

\bibitem[\protect\citeauthoryear{Videler}{1977}]{videler1977mechanical}
J.~Videler.
\newblock Mechanical properties of fish tail joints.
\newblock {\em Fortschritte der Zoologie}, 24(2-3):183--194, 1977.

\bibitem[\protect\citeauthoryear{Webb}{}]{webb1994biology}
P.~Webb.
\newblock {\em {The biology of swimming}}, pages 45--62.
\newblock Cambridge University Press, Cambridge, UK.

\bibitem[\protect\citeauthoryear{Webb and Cotel}{2010}]{webb2010turbulence}
P.~Webb and A.~Cotel.
\newblock {Turbulence: does vorticity affect the structure and shape of body
  and fin propulsors?}
\newblock {\em Integrative and Comparative Biology}, page icq020, 2010.

\bibitem[\protect\citeauthoryear{Wenger \bgroup \em et al.\egroup
  }{2007}]{wenger2007mechanical}
M.~P. Wenger, L.~Bozec, M.~A. Horton, and P.~Mesquida.
\newblock Mechanical properties of collagen fibrils.
\newblock {\em Biophysical Journal}, 93(4):1255--1263, 2007.

\bibitem[\protect\citeauthoryear{Westneat \bgroup \em et al.\egroup
  }{2004}]{westneat2004structure}
M.~W. Westneat, D.~H. Thorsen, J.~A. Walker, and M.~E. Hale.
\newblock {Structure, function, and neural control of pectoral fins in fishes}.
\newblock {\em IEEE Journal of Oceanic Engineering}, 29(3):674--683, 2004.

\bibitem[\protect\citeauthoryear{Zhu and Shoele}{2008}]{zhu2008propulsion}
Q.~Zhu and K.~Shoele.
\newblock {Propulsion performance of a skeleton-strengthened fin}.
\newblock {\em Journal of Experimental Biology}, 211(13):2087--2100, 2008.

\end{thebibliography}

\section*{Appendix A -- $N$-ray fin}

With few modifications, the bending-stretching coupling formalism of the 2-ray fin applies equally well to the $N$-ray fin.
The key difference is that we decompose the $N$-ray fin into $N$ independent Fourier modes and analyse each mode separately.
The dependence of stiffness for each Fourier mode on curvature are similar to each other and to the 2-ray stiffness.

\begin{figure}[ht]
{\includegraphics{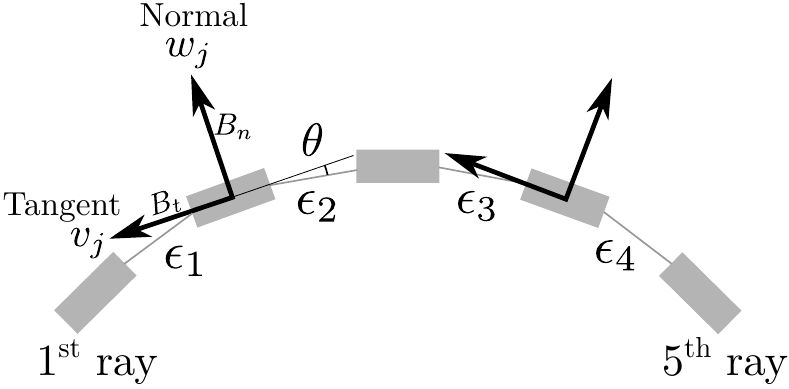}}
\caption{A $N$-ray fin with $N$ rays and $N-1$ membranes. In this example, $N = 5$. The tangent axis of each ray forms an angle $\theta$ with neighbouring membranes and the rays trace out a constant transverse curvature. 
Each ray has a set of normal and tangent axis with bending values $w_j(x)$ and $v_j(x)$, respectively, for the $j^\text{th}$ ray and the elongation for the $j^\text{th}$ membrane is $\str_j(x)$. (c) The part of the membranes $AA'$ and $BB'$ on which the free-body diagram is considered. The membranes attach to the rays at the points $A'$ and $B'$ respectively.
}
\label{fig:nray}
\end{figure}

Fin curvature mediates the coupling between the bending of rays and stretching of membranes of the $N$-ray fin in the same manner as the 2-ray fin.
Figure \ref{fig:nray} shows a cross-section of the $N$-ray fin at a distance $x$ from the fin base.
The bending-stretching coupling of the $N$-ray fin appears as a geometric relation for every pair of adjacent rays, which satisfies
\begin{align}
&\str_j = (w_j + w_{j+1})\sin\theta + (v_j - v_{j+1})\cos\theta, && \qquad\qquad\qquad\qquad \ \ \ \text{for } j = 1,2, \dots, N-1
\label{eqn:nraygeo}
\end{align}
where $w_j(x)$ and $v_j(x)$ are the displacements of the $j^\text{th}$ ray along the normal and tangent axes, respectively, and $\str_j(x)$ is the elongation of the $j^\text{th}$ membrane.
The coupling also appears in the balance of ray resistance to bending and membrane springy distributed forces for each ray, namely,
\begin{align}
&\Bn w''''_j = -k(\str_{j-1} + \str_j)\sin\theta, \quad \Bt v''''_j = -k(\str_{j-1}-\str_j)\cos\theta, && \text{for } j = 1,2, \dots, N
\label{eqn:nray}
\end{align}
where nonexistent $0^\text{th}$ and $N^\text{th}$ membranes have zero elongation ($\str_0 = \str_N = 0$). 
Each ray is clamped at the base and is loaded at the ray tips by $f_j$ along the normal direction as
\begin{gather}
\begin{aligned}
&w_j= w'_j = v_j = v'_j = 0							 &\text{ at } x = 0\ \\
&w''_j = v''_j = \Bt v'''_j = 0, \quad \Bn w'''_j = f_j    &\text{ at } x = L.\\
\end{aligned}%
\label{eqn:nraybc}
\end{gather}

A Fourier decomposition of the $N$-ray fin transforms the coupling mechanics into equations that are similar to the 2-ray fin.
The decomposition trades the displacements $w_j(x)$ and $v_j(x)$, elongations $\str_j(x)$, and loads $f_j$ for their corresponding Fourier coefficients 
$\what_m(x)$, $\vhat_m(x)$, $\shat_m(x)$, and $\fhat_m(x)$, respectively, and satisfies
\begin{gather}
\begin{aligned}
&\str_j = \sum_{m=1}^{N} \shat_m \sin \left(\dfrac{\pi mj}{N}\right), 
&w_j = \sum_{m=1}^{N}    \what_m \sin \left(\dfrac{\pi m (j-1/2)}{N}\right), \\
&v_j = \sum_{m=1}^{N}    \vhat_m \cos \left(\dfrac{\pi m (j-1/2)}{N}\right), 
&f_j = \sum_{m=1}^{N}    \fhat_m \sin \left(\dfrac{\pi m (j-1/2)}{N}\right). 
\label{eqn:DFT}
\end{aligned}
\end{gather}
Under this decomposition, the geometric relation for each pair of adjacent rays described in equation \eqref{eqn:nraygeo} transforms to
\begin{align}
\shat_m = 2\what_m s_m + 2\vhat_m c_m, \quad \text{ where} \quad s_m = \sin\theta\cos\left(\dfrac{m\pi}{2N}\right), \quad c_m = \cos\theta\sin\left(\dfrac{m\pi}{2N}\right).
\label{eqn:Fouriergeo}
\end{align}
Likewise, the balance of forces on the $j^\text{th}$ ray in equation \eqref{eqn:nray} becomes
\begin{align}
\Bn\what''''_m = -2k\shat_m s_m, \quad \Bt\vhat''''_m = -2k\shat_m c_m,
\label{eqn:Fourier}
\end{align}
and the clamping at the base and loading at the tip of equation \eqref{eqn:nraybc} becomes
\begin{gather}
\begin{aligned}
&\what_m= \what'_m = \vhat_m = \vhat'_m = 0							 &\text{ at } x = 0\ \\
&\what''_m = \vhat''_m = \Bt \vhat'''_m = 0, \quad \Bn \what'''_m = \fhat_m    &\text{ at } x = L.\\
\end{aligned}%
\end{gather}
Furthermore, a physical length scale $l_m$ and ray anisotropy parameter $\gamma_m$ can be extracted for each Fourier mode as
\begin{align}
\shat''''_m = -\dfrac{1}{l_m^4}\shat_m, \quad \text{ where }\quad \dfrac{1}{l_m^4} = \dfrac{4kc^2_m}{\Bt}\left( 1+ \gamma_m \right),\quad \gamma_m = \dfrac{\Bt}{\Bn}\dfrac{s^2_m}{c^2_m}.
\label{eqn:Fourierstr}
\end{align}
The multiplicative constants differ but the Fourier mode mechanics in equations (\ref{eqn:Fouriergeo} - \ref{eqn:Fourierstr}) mirror the 2-ray fin mechanics. Consequently, the Fourier modes can be understood as a set of 2-ray fins with separate length scales $l_m$ and ray anisotropy parameters $\gamma_m$.

The curvature-induced stiffening of a $N$-ray fin is an emergent property of the collective sum of its sinusoidal modes, and is fundamentally a result of a competition between ray bending and membrane stretching.
This competition appears in similar form for the 2-ray fin and for all the modes.
Although stiffening in the $N$-ray fin is a convoluted result of the competition in each mode, it retains the same qualitative behaviour as a 2-ray fin. 
That is to say, the $N$-ray fin can be similarly understood as a combined effect of normal bending, and tangential bending and membrane stretching, but with contributions from multiple rays and membranes. 

To estimate the range of stiffnesses spanned by the fin, let us consider the simplification of a single Fourier mode, $m=1$.
In this case, the angle $\theta$ ranges from 0 for a flat fin to $\pi/2N$ for a quarter-cylindrical fin (to maintain analogy with results in (\ref{eqn:Loverlmax}-\ref{eqn:findesigncriteria})). 
The expression for $l_1$ then simplifies to
\begin{align}
 \dfrac{1}{l_1^4} = 4k \left( \dfrac{c_1^2}{\Bt} + \dfrac{s_1^2}{\Bn} \right).
\end{align}
This corresponds to maximum value of $L/l_1$ governed by
\begin{align}
\left(\dfrac{L}{l_1}\right)_\text{max} = k^{1/4} \sqrt{\sin (\pi/N)} \left( \dfrac{1}{\Bt} + \dfrac{1}{\Bn}\right)^{1/4}.
\label{eqn:nrayloverl}
\end{align}
In the limit, $N\gg 1$ and $\Bt \gg \Bn$, this expression simplifies to
\begin{align}
\left(\dfrac{L}{l_1}\right)_\text{max} = \sqrt{\dfrac{\pi}{N}} \left(\dfrac{k}{\Bn}\right)^{1/4}.
\label{eqn:nrayloverlasymp}
\end{align}

Similarly, $\gamma_1$ spans the range
\begin{align}
 0 \le \gamma_1 \le \dfrac{\Bt}{\Bn} \dfrac{\sin^2 (\pi/2N) \cos^2(\pi/2N) }{ \cos^2(\pi/2N) \sin^2(\pi/2N) } = \dfrac{\Bt}{\Bn}.
 \label{eqn:nraygamma}
\end{align}
Equation \eqref{eqn:nrayloverl} has an extra factor of $\sqrt{\sin(\pi/N)}$ for $L/l_1$ that accounts for the presence of $N$ rays, compared to $L/l$ in \eqref{eqn:Loverlmax} for the 2-ray analysis, while the upper bound on $\gamma$ in \eqref{eqn:nraygamma} agrees with that in \eqref{eqn:Loverlmax}.
It can be easily verified that using a half-cylinder instead of a quarter-cylinder does not change the scaling of the bounds on the stiffness of the $N$-ray fin.
But the analogous substitution $\theta=\pi/2$ in the 2-ray analysis yields an unbounded value of $\gamma$, which is not representative of the $N$-ray fin. 
Therefore, to maintain analogy with the $N$-ray fin, we use $\pi/4$ as a representative value of $\theta$ in the 2-ray fin analysis.
Furthermore, imposing the condition that $L/l_1 \sim \gamma^{1/3}$ in (\ref{eqn:nrayloverlasymp}-\ref{eqn:nraygamma}), leads to \eqref{eqn:findesigncriteria}.

\section*{Appendix B -- Bulging of membrane}
We finally show that the out-of-plane bulging of the membrane, if present, merely influences the parameters we substitute in the model, but does not change the model itself.
In our model, we have assumed the external force to be exerted on the rays, but the membrane presented at least as much area to the surrounding fluid as the rays, if not more.
In addition, we assumed that the membrane deformation is restricted to stretching in the plane, but in practice the membrane could also bulge out of plane.
Here we show that the hydrodynamic pressure acting on the membranes is transmitted to the adjoining rays through a bulging deformation of the membrane.

\begin{figure}
\includegraphics{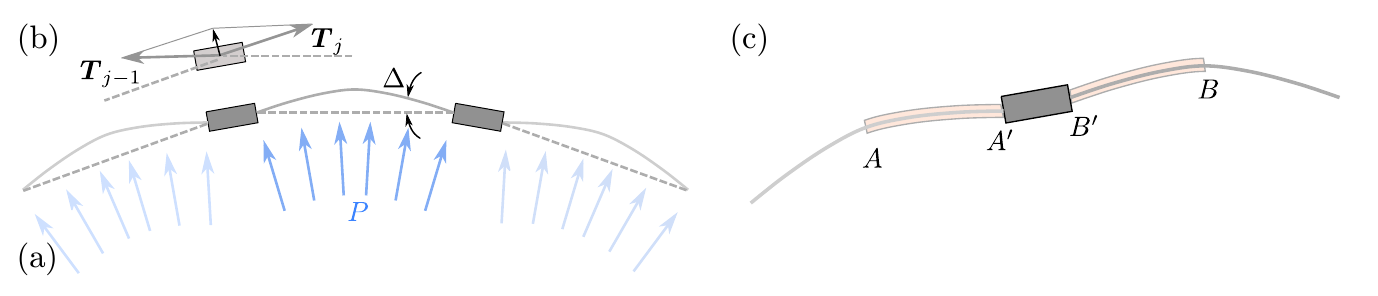}
\caption{Schematic used to estimate the influence of membrane bulging. (a) The cross section of rays (shaded rectangles) and membranes (curves) is externally loaded by a hydrodynamic pressure difference across the membrane surface, which is labelled $P$ (arrows). This loading causes the membranes to bulge out of plane and a tension $\boldsymbol{T}$ to develop in them. (b) The resultant of the tensions from the membranes on the two sides of a ray is, to leading order, perpendicular to the fin. (c) Shaded rectangles between $AA'$ and $BB'$ show control volume for constructing a free body diagram. }
\label{fig:BulgeSchematic}
\end{figure}

Consider the schematic shown in figure~\ref{fig:BulgeSchematic} of a cross section of two adjacent rays and the connected membranes, to model the influence of the hydrodynamic pressure difference acting on the membranes.
At this cross section, the pressure causes the $j^\text{th}$ membrane to bear a tension $\bT_j$ aligned transverse to the rays, and to bulge with curvature $\kappa_j$.
The pressure difference is related to the tension and the curvature by the Young-Laplace equation as 
\begin{align}
 P = \kappa |\bT|.
\end{align}

The tension in the membrane is established due to the rays being pulled apart by the fish's musculature, in which case we consider the tension to be an externally imposed parameter.
If the magnitude of the tension is much greater than $P S$ ($S$ is the width of the membrane), then $\kappa = P/|\bT| \ll P/PS = 1/S$.
In this case, the out-of-plane deformation due to bulging $\kappa S^2 \ll S$, and therefore the bulging causes a negligible change in the overall shape of the membrane, except for one factor.
The bulging changes the local orientation of the membrane where it attaches to the ray by a small amount $\Delta = \sin^{-1} \left(\kappa S/2\right) = \sin^{-1} \left(PS/2|\bT| \right)$.
This slight redirection of the membrane tensions  $\bT_{j-1}$ and $\bT_j$ on both sides of a ray lead to an unbalanced force, as shown in Figure~\ref{fig:BulgeSchematic}(b), of $2 |\bT| \sin \Delta = PS$ perpendicular to the fin surface.
Note that $PS$ is merely the force of hydrodynamic pressure $P$ acting on membrane of length $S$, which is being transmitted to the rays.
Parallel to the fin surface the membrane tension from the opposite sides cancel out to leading order.
It is around this base state that our analysis in \S\ref{sec:mathmodel} and in Appendix A applies.
A further differential in the displacement of the rays that stretch the membrane contribute to the unbalanced force, as written in \eqref{eqn:nraygeo}.

While our analysis in \S \ref{sec:mathmodel} directly treats the case of a taut membrane and a small bulge, an essentially identical model results in the general case in which the  membrane may be slack.
In the general case, consider the free body diagram for half the membranes on either side of a ray, as shown by the shaded region $AA'$ and $BB'$ in Figure~\ref{fig:BulgeSchematic}(c). 
The hydrodynamic force on the membrane is balanced by the force exerted by the ray at the points $A'$ and $B'$ of attachment to the membrane.
This result follows independent of the tension in the membrane and its deformation.
The unbalanced force on the rays perpendicular to the fin surface that results from the membrane attachment may in this manner be ultimately traced to the hydrodynamic pressure force on the membrane.
The fictitious state where the rays are undeformed but the membrane bulges out in response to the hydrodynamic pressure, is the reference state about which we consider the deformation of the rays.
Deformations about this state essentially leads to the model described by equations \eqref{eqn:nraygeo}, \eqref{eqn:nray}, and \eqref{eqn:nraybc}.

As the rays deform about this state, the tension in the membranes and the angle at the points of attachment change. 
The unbalanced force on the rays changes as the rays deform; the change in the unbalanced force to leading order is proportional to these deformations, which themselves are proportional to the amount of bending of the rays.
The proportionality constants are embodied in a lumped parameter, which we represent by $k$, the membrane stiffness.
The case of a slack membrane may be considered to be equivalent to the limit of vanishingly small $k$ in the linear response regime.
In this manner, our model provides a useful approximation in the general case whether or not the membrane is stretched by the action of musculature connected to the fish fin.
As described earlier, we considered the external loading on the fin to be applied at the tips of the rays to mimic experimental measurements of the fin stiffness; however, applying the external force on the rays instead of the membranes is not a limitation of our analysis.
Furthermore, the response of the fin to a point load is essentially a Green's function for determining the deformation under distributed loading.
In this manner, our solution provides the foundation for developing and implementing a method that relates the distributed hydrodynamic pressure loading to the deformation of the fin structure using a Green's function.

\end{document}